\documentclass[11pt,letterpaper]{article}

\usepackage[margin=1in]{geometry}
\usepackage[T1]{fontenc}
\usepackage[utf8]{inputenc}
\usepackage{lmodern}
\usepackage{graphicx}
\usepackage{booktabs}
\usepackage{colortbl}
\usepackage{amsmath}
\usepackage{amssymb}
\usepackage[table]{xcolor}
\usepackage{enumitem}
\usepackage{url}
\usepackage{float}
\usepackage[hidelinks]{hyperref}

\setlist{topsep=2pt,itemsep=1pt,parsep=0pt}
\newcommand{\code}[1]{\texttt{\small #1}
} 

\title{\textbf{Composable CXL Memory as a Kubernetes-Native\\
Shared Memory for LLM Serving}
\\[4pt] \large A Feasibility Study on Seagate Composable Memory Appliance}

\author{Hongjian Fan, Kevin Zhang, David Habinsky, Matthew Carper and Sean
Dykstra\\
\textit{Research Group, Seagate Technology, LLC}}

\begin{document}
\date{}
\maketitle

\begin{abstract}
We present a Kubernetes Dynamic Resource Allocation (DRA) driver that makes
composable Compute Express Link (CXL) memory a schedulable cluster resource, and
evaluate the resulting shared-memory tier for cross-node KV-cache transfer and
reuse in LLM serving. The driver composes CXL regions on demand, materializes
them as DAX devices on each participating host, and injects them into pods under
a single Container Device Interface (CDI) name. A shared-memory connector for
vLLM uses that region as a KV-cache tier with a slot directory embedded inside
the shared medium, which requires no external metadata service. On a two-node
cluster with a 512\,GiB CXL appliance and Qwen2.5-7B-Instruct, cross-node prefix
reuse reduces TTFT by 5.5$\times$--36.6$\times$ at an external hit rate of
95.4--99.5\,\%, while node-local tiers (GPU prefix caching, CPU-DRAM offload)
fall back to full recompute. The sharing gap, defined as the latency ratio
between cross-node and same-node reuse, is 1--4\%, indicating that cross-node
reuse incurs little additional latency relative to same-node reuse. On a KV
cache transfer case between disaggregated prefill/decode, CXL and RDMA converge
to within 7\,\% at all prefix lengths. We report these results as a feasibility
study, not a performance evaluation.
\end{abstract}

\section{Introduction}

The KV cache makes autoregressive serving fast and expensive. For
Qwen2.5-7B-Instruct, every token in the context takes 56\,KiB of key
and value tensors, so a 32\,K-token prompt is 1.75\,GiB of KV cache.
Long shared prefixes are common in multi-turn chat,
retrieval-augmented generation (RAG), and agentic workflows. Reusing
the KV cache lets later requests skip recomputation. Modern engines
exploit this aggressively with prefix caching and KV
offload~\cite{vllm,sglang,lmcache,cacheblend}, and with
prefill/decode (P/D) disaggregation~\cite{distserve,splitwise,mooncake}.

What this needs is a cache tier that is both \emph{large} and
\emph{shared}. The tiers available today provide one property or the
other (Table~\ref{tab:configs}):

\begin{itemize}
\item \textbf{T1: GPU VRAM prefix cache}~\cite{vllm,sglang}, fastest,
  but smallest and confined to a single GPU.
\item \textbf{T2: CPU DRAM offload}~\cite{lmcache,cacheblend}, an order
  of magnitude larger, but confined to a single node.
\item \textbf{T3: pooled RDMA store} (Mooncake Store over
  RoCEv2~\cite{mooncake,nixl}), shared, but every reuse traverses a
  transfer protocol and a NIC.
\end{itemize}

A request that misses every tier falls back to \textbf{T0}, the
recompute floor.

\paragraph{Serving on Kubernetes.} LLM serving is deployed the way
most production workloads are: as pods on Kubernetes. Every request
first goes through a compute-bound \emph{prefill} and then a
memory-bound \emph{decode}~\cite{llm-survey};
P/D disaggregation runs these phases on different nodes. This makes KV
reuse a cross-node problem: a prefix computed on one node is reused by
a request on the other instead of being recomputed.

\begin{figure}[t!]\centering
\includegraphics[width=0.78\linewidth]{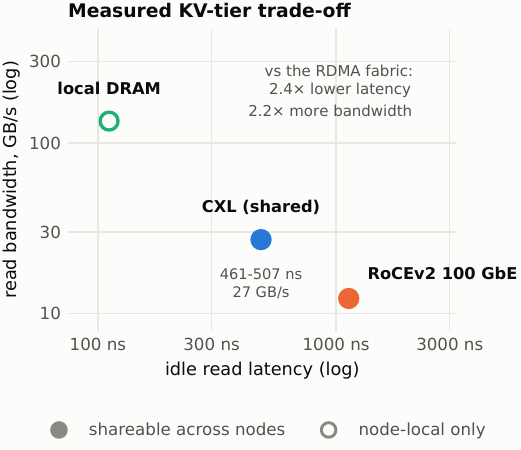}
\caption{Measured trade-off between the candidate KV-reuse tiers on our
testbed. T2 (CPU DRAM) is fastest but node-private; the RoCEv2 fabric
carrying T3 is shared but protocol-bound. CXL (T4) is the only
shareable tier within $\sim$4.5$\times$ of local-DRAM latency, using
ordinary loads and offering pooled capacity beyond per-node budgets.}
\label{fig:tiers}
\end{figure}

\paragraph{The fourth tier.} Compute Express Link
(CXL)~\cite{cxlspec} offers the missing tier, \textbf{T4}. A CXL
\emph{composable memory appliance}~\cite{cma,cma_ocp_talk} offers a pool of
memory blades behind several host ports. The appliance carves out a
region and assigns that region to several hosts at once. On each host
the region appears as ordinary byte-addressable memory, reachable at
sub-microsecond latency. On our testbed the region reads at
27--28\,GB/s, 2.2$\times$ the 12.25\,GB/s of the RoCEv2 fabric, at
2.4$\times$ lower latency than the RDMA tier (Figure~\ref{fig:tiers}).
That sits within the 2--5$\times$ local-DRAM latency reported for CXL
platforms~\cite{tpp,demystify,hitchhikers-cxl}. And because the region
is accessed with plain load/store memory semantics, there is no
transport layer on the reuse path. A KV block loads with a
\code{memcpy} from a shared physical address.

\paragraph{The gap.} No Kubernetes resource type captures a \emph{schedulable
shared-memory resource}. The Pangaea\,v2 project~\cite{pangaea2} brings CXL
memory into Kubernetes via an NRI plugin but only on a per-host basis. Recent
CXL-KV systems~\cite{tract,hymcache} assume the region already exists. Without a
schedulable resource, shared CXL memory cannot be multiplexed between tenants or
released on claim deletion. Dynamic Resource Allocation
(DRA)~\cite{drakep,dra-arch} (\code{resource.k8s.io/v1} since v1.34) generalizes
the device-plugin model used by node-local accelerators such as GPUs, NICs, and
FPGAs. Kubernetes cannot give a pod a device that does not yet exist, or one
attached to more than one node. Closing this gap requires extending DRA: a
driver must call the appliance's fabric manager to compose a region on demand
and inject it into pods on every participating node.

\paragraph{This paper.} A DRA plugin for the
Composability Fabric Manager (CFM) of the Seagate composable memory
appliance makes its memory a schedulable Kubernetes resource. We use
the resulting region as a shared-memory tier (T4) for LLM serving
(vLLM inside llm-d~\cite{llmd}). Our contributions:

\begin{enumerate}
\item \textbf{A Kubernetes DRA driver for composable CXL memory}
  (\S\ref{sec:dra}). It advertises blade capacity as a \code{ResourceSlice} whose
  \code{NodeSelector} spans every host cabled to the blade; composes a region on
  demand through a fabric manager; materializes it as a DAX sub-device on
  \emph{each} participating host; and injects it under a single Container Device
  Interface (CDI)~\cite{cdi} name across all nodes. A finalizer and redundant
  teardown metadata make free-on-delete survive a plugin restart. \item \textbf{An
  in-region metadata shared-memory KV tier} (\S\ref{sec:connector}). The directory
  resides inside the shared medium itself with an atomic slot bitmap plus a
  parallel key array, so presence lookup requires no external metadata service.
  \item \textbf{An evaluation of shared memory for cross-node KV reuse}
  (\S\ref{sec:eval}). A two-node cluster, a 512\,GiB appliance mapped to both
  hosts, and a cross-replica reuse experiment designed so that \emph{only} a
  shared tier can produce a hit. We report the headline result, the \emph{sharing
  gap}, together with the costs: the local premium, the achieved fraction of
  fabric bandwidth, and an assessment of data integrity under cross-node access.
\end{enumerate}

\paragraph{Scope.} This is a feasibility report. During the KV cache reuse
study, both replicas are full engines. We also tested KV cache transfer,
with the system disaggregated into prefill and decode over both CXL
and RDMA. The harness is closed-loop and runs one
session at a time, so time-to-first-token (TTFT) is the metric it supports;
we make no goodput or tail-latency claim. Zero-copy GPU$\leftrightarrow$CXL
DMA and a multi-tenant scheduling experiment are explicit non-goals.
\S\ref{sec:limits} enumerates the remaining limitations.

\section{Scheduling Composable CXL Memory}
\label{sec:dra}

\begin{figure}[t!]\centering
\includegraphics[width=0.92\linewidth]{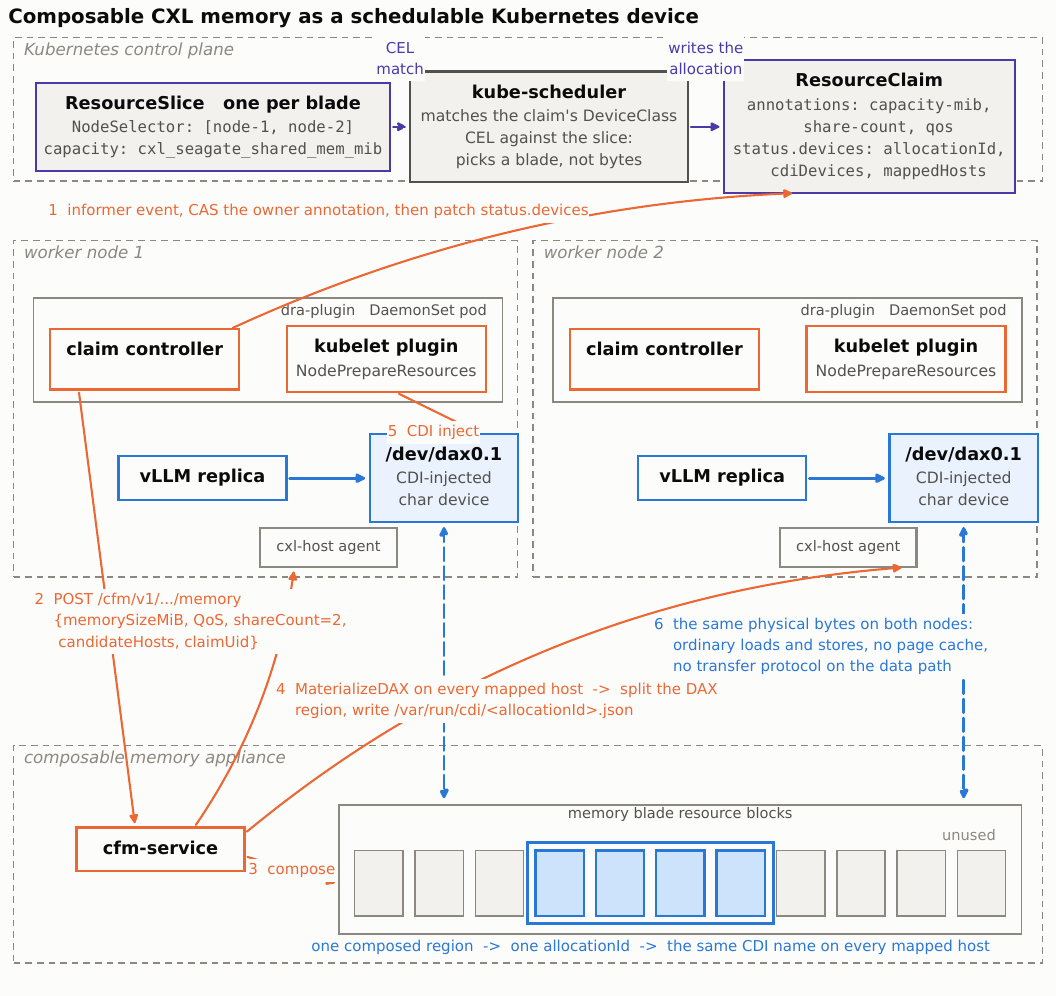}
\caption{The DRA driver in context. Control-plane objects on the left: a
\code{ResourceSlice} per blade whose node selector spans the connected hosts,
and a \code{ResourceClaim} carrying size, share count and QoS in
\emph{annotations}. The \code{dra-plugin} DaemonSet on each node runs both a
claim controller (one of which is elected to compose) and a kubelet plugin.
\code{cfm-service} drives the blade over Redfish; \code{cxl-host} materializes
the region locally. The result is one composed region reaching a pod on each
node under a single CDI name.}
\label{fig:dra}
\end{figure}

Our driver, \code{cfm-dra-plugin.cxl.seagate.com}, runs as a DaemonSet on every
CXL-capable node (Figure~\ref{fig:dra}).

\subsection{Resource model}

The driver publishes \textbf{one \code{ResourceSlice} per blade}, containing a
single device with attributes \code{applianceId} and \code{bladeId} and one
capacity, \code{cxl\_seagate\_shared\_mem\_mib}, computed from the blade's
\emph{unallocated} resource blocks.

The key distinction is the slice's node scope. A node-local driver sets
\code{NodeName}; we set a \textbf{\code{NodeSelector}} matching
\code{kubernetes.io/hostname} over exactly the hosts cabled to that blade. The
scheduler may therefore place a consumer on any of them, since the region will
be reachable from all. The cost is that the scheduler cannot express a locality
preference finer than ``connected,'' and the driver---not the
scheduler---decides which ports are assigned.

A cluster-scoped \code{DeviceClass} selects the driver with
\code{device.driver == "cfm-dra-plugin.cxl.seagate.com"}; a workload's claim
requests \code{deviceClassName: cfm-shared-memory}.

\subsection{Region sizing via annotations}
\label{sec:sizing}

\begin{figure}[t!]\centering
\includegraphics[width=0.72\linewidth]{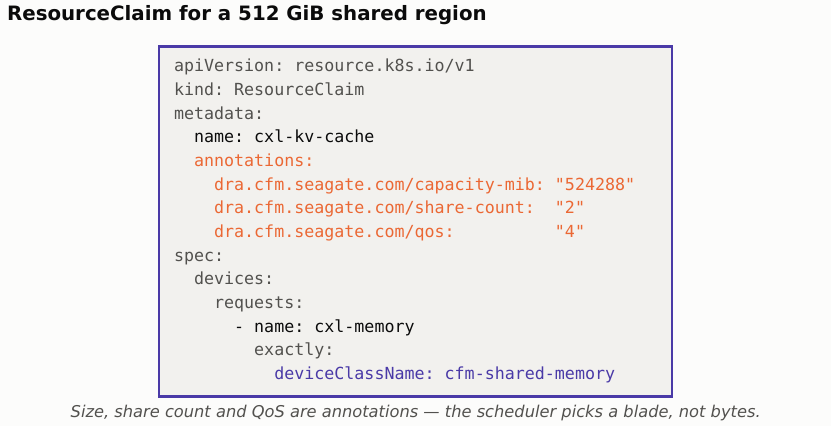}
\caption{A claim for a 512\,GiB region shared by two nodes. Size, share count
and QoS are annotations, not part of the device request
(\S\ref{sec:sizing}). Both vLLM pods reference this one claim, so both access
the same composed region.}
\label{fig:claim}
\end{figure}

A DRA \code{DeviceRequest} carries a device class and Common Expression Language
(CEL) selectors. It does not carry a quantity. Region size, share count and QoS
class therefore travel as \textbf{annotations on the \code{ResourceClaim}}
(\code{dra.cfm.seagate.com/capacity-mib}, \code{/share-count}, \code{/qos}),
read by our controller after allocation. Figure~\ref{fig:claim} shows a complete
claim.

The consequence is that the scheduler picks a \emph{blade}, never a byte count.
The advertised capacity is not decremented when a claim is allocated; it shrinks
only on the next 30-second publish cycle. There is therefore no scheduler-side
capacity accounting or overcommit protection: two claims created close together
can be admitted against the same free capacity, and the second fails at compose
time.

\subsection{Compose-on-demand}

\begin{figure}[H]\centering
\includegraphics[width=0.92\linewidth]{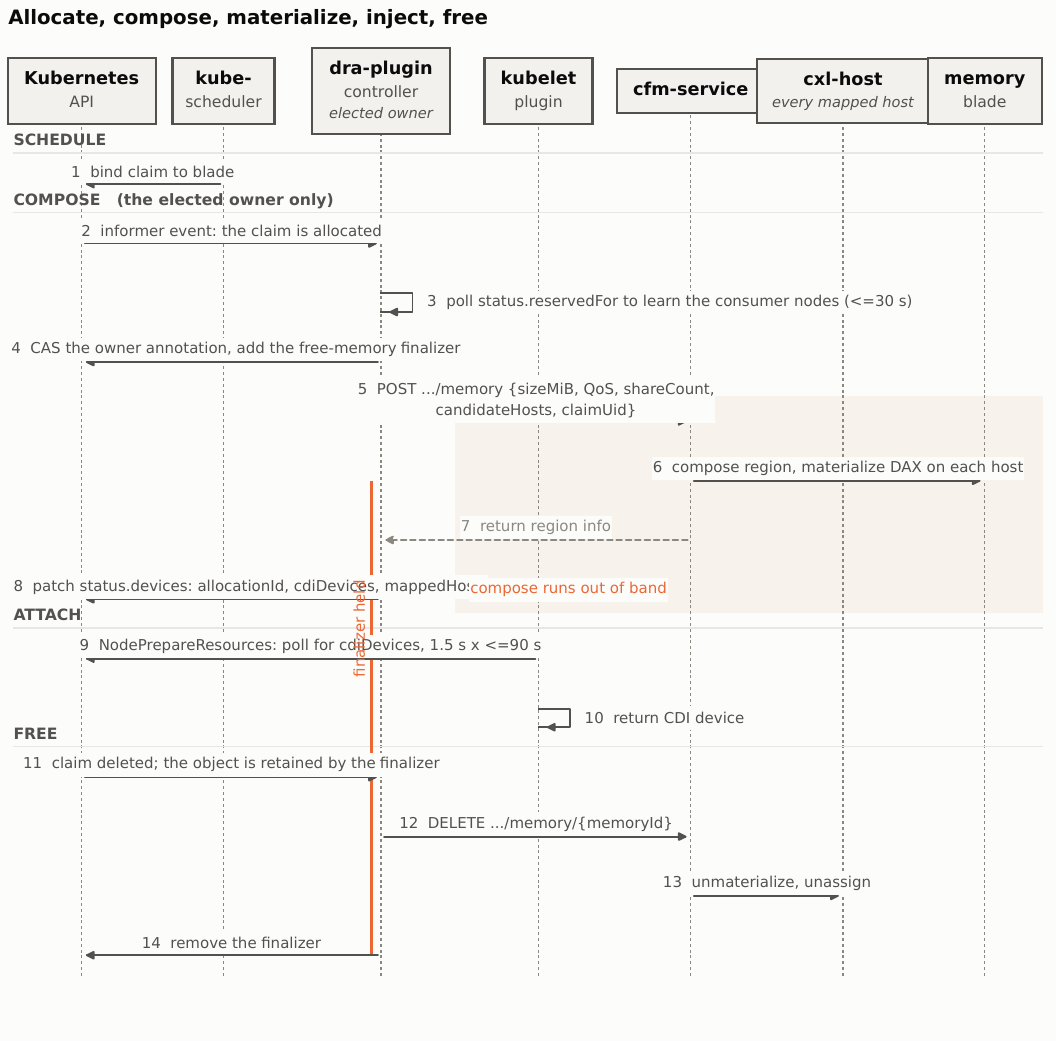}
\caption{The
allocate$\rightarrow$compose$\rightarrow$materialize$\rightarrow$inject$\rightarrow$free
sequence. Note the owner election among DaemonSet replicas, the span over which
the finalizer is held, and the shaded out-of-band window: the region is composed
by a controller, not by the node that kubelet asks to prepare the claim, so
\code{PrepareResourceClaims} has to wait for CDI names to appear in the claim
status.}
\label{fig:sequence}
\end{figure}

Figure~\ref{fig:sequence} traces the lifecycle. When the scheduler allocates
the claim, every DaemonSet replica sees it:

\begin{enumerate}
\item \textbf{Wait for a consumer.} The controller waits (up to 30\,s) for a pod
  referencing the claim to be bound to a node. \item \textbf{Elect an owner.}
  Candidates race a compare-and-set on the \code{dra.cfm.seagate.com/owner}
  annotation; exactly one wins and calls the fabric manager. Standard DRA drivers
  need no election because they run as a single controller or work node-locally.
  \item \textbf{Take a finalizer.} The winner adds
  \code{dra.cfm.seagate.com/free-memory} before composing anything, so a claim
  deleted mid-flight cannot vanish before blade capacity is released. \item
  \textbf{Compose.} It reads capacity, share count and QoS from the annotations
  (Figure~\ref{fig:claim}), builds the candidate host list from the blade's port
  topology, and issues \code{ComposeMemory} to \code{cfm-service}. The blade
  composes a region from resource blocks matching the QoS class and
  \emph{multi-assigns} it to each selected port. \item \textbf{Materialize per
  host.} For each assigned port, \code{cfm-service} invokes a Redfish action on
  the connected node's \code{cxl-host} daemon, which creates the local DAX device
  and writes a CDI spec (\S\ref{sec:materialize}). \item \textbf{Publish.} The
  controller patches the resulting CDI device names into
  \code{ResourceClaim.Status.Devices}, which kubelet reads.
\end{enumerate}

\subsection{Single CDI name across nodes}

The same region, identified by one \code{allocationId}, is materialized
independently on each host, possibly at different DAX indices, since numbering
is local. The CDI device name, however, is the allocation UUID, identical on
every node. A pod spec that names one claim therefore resolves, on whichever
node it lands, to a node-local character device pointing at the composed region.

\subsection{Host-side materialization and injection}
\label{sec:materialize}

\code{cxl-host} runs privileged on each CXL node and exposes a Redfish API to
the fabric manager. On a materialize action, it:

\begin{enumerate}
\item For CXL 1.1/2.0 devices, calls the \textbf{sysfs DAX split}: writing to
  \code{/sys/bus/dax/devices/} to carve a sub-device of the requested size,
  preceded if necessary by a pad device to align the offset. For CXL 3.x devices
  with DCD (dynamic capacity device) support, it requests capacity extension
  directly from the CXL kernel driver. \item Writes a CDI spec to
  \code{/var/run/cdi/<allocationId>.json} of kind \code{cxl.seagate.com/dax},
  containing the device node to bind-mount plus the environment the workload
  consumes: \code{CXL\_DAX\_DEVICE} and \code{CXL\_DAX\_SIZE}.
\end{enumerate}

Carving a sub-device per allocation gives per-allocation isolation inside a
physically shared pool.

The kubelet plugin's \code{PrepareResourceClaims} reads CDI device IDs from the
claim status, polling for up to 90\,s because composition happens out of
band relative to kubelet's request (the shaded window in
Figure~\ref{fig:sequence}). \code{Unprepare} is a no-op: memory is released by
the controller on claim deletion.

\subsection{Lifecycle and teardown}

The driver persists teardown metadata in three places: in memory, as a claim
annotation, and in the claim's device status. It reconstructs free parameters
from whichever copy survives. Combined with the finalizer, this ensures
free-on-delete works across a plugin restart. If \code{FreeMemory} keeps
failing, the controller retries five times and then removes the finalizer
anyway, choosing a possible capacity leak over a claim that can never be
deleted; the leak is visible in the fabric manager and recoverable out of band.

Two failure behaviors are deliberate. Materialization is \emph{non-fatal}: if
one host fails to create its DAX device, the compose still succeeds and the pod
on that node starts without a device. And if a blade is power-cycled, the fabric
manager does not rediscover state automatically, because the current prototype
does not persist composed-region metadata across power cycles. An operator must
resynchronize the state after a power cycle.

The driver is written in GoLang against \code{resource.k8s.io/v1} and the
\code{k8s.io/dynamic-resource-allocation} kubelet-plugin framework; it requires
Kubernetes v1.34+ for GA DRA.

\section{A Shared Memory Backend for LLM Serving}
\label{sec:connector}

\begin{figure}[t!]\centering
\includegraphics[width=0.95\linewidth]{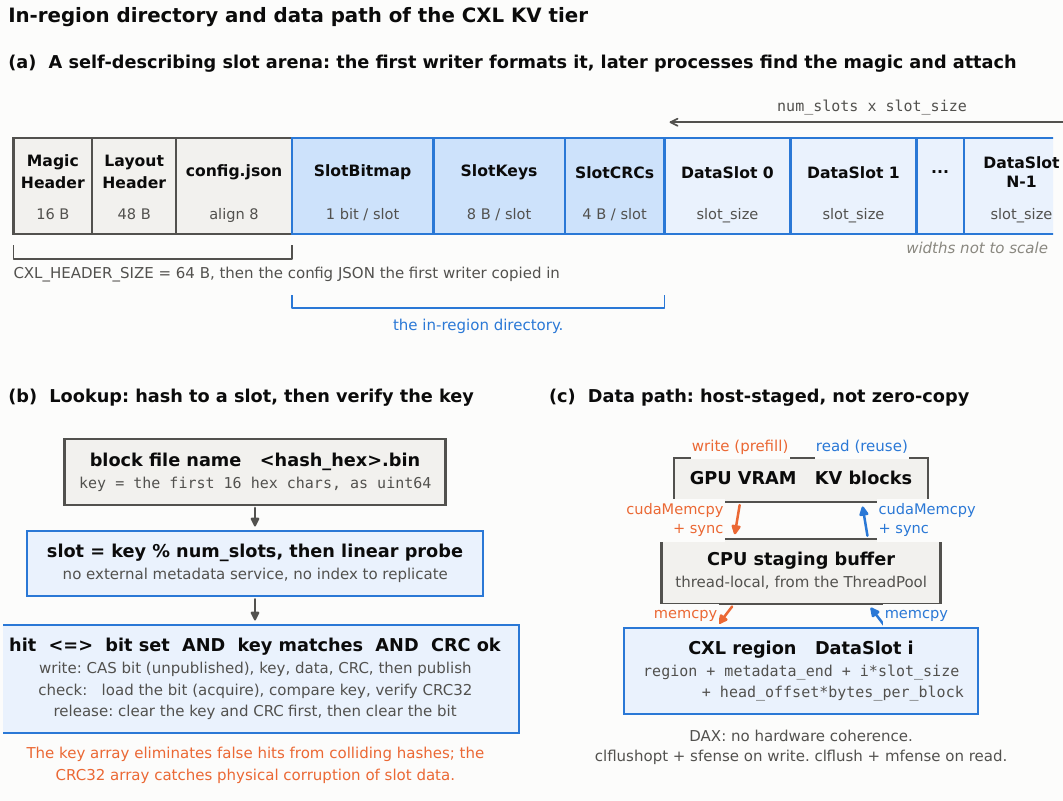}
\caption{The shared region. (a) The self-describing layout: magic and layout
headers, the writer's configuration, the slot bitmap, the parallel key array,
and the per-slot CRC32 array, then fixed-size data slots. (b) The lookup
protocol: a slot key derived from the block-hash filename,
\code{key \% num\_slots} with linear probing, and a hit only when the bit is set
\emph{and} the stored key matches, with the acquire/release ordering that makes
it safe across processes. (c) The host-staged data path and the coherence rules
for DAX versus file-backed mappings.}
\label{fig:layout}
\end{figure}

The connector is a storage backend behind vLLM's \code{OffloadingConnector}: a
control-plane manager in the scheduler process answers lookups and reserves
slots, and per-worker handlers move tensors. The device path and size come from
the CDI environment injected in \S\ref{sec:materialize}, so the same deployment
manifest works on any node.

\subsection{Mapping the region}

A C++ engine opens the DAX device and \code{mmap}s it \code{MAP\_SHARED}, then
calls \code{madvise(MADV\_RANDOM)} to suppress readahead. The optional
\code{offset} argument enables multi-tenant partitioning: two independent pools
can live in one device at different offsets. Because DAX has no page cache, the
mapping \emph{directly references} the device memory.

\subsection{A self-describing region}

Figure~\ref{fig:layout}(a) shows the layout:

\begin{center}\small
\code{MagicHeader | LayoutHeader | config.json | SlotBitmap | SlotKeys | SlotCRCs | DataSlot\dots}
\end{center}

The magic header signals whether the region is formatted. The layout header
records slot size, slot count, and the end of the in-region directory area; the
writer's \code{config.json} (model, dtype, block sizes) lets a later attacher
validate compatibility. Slot size is derived from the engine's block geometry,
and slot count from the region size.

The first process to map an unformatted region initializes it (headers,
configuration, zeroed bitmap). Any later process \emph{attaches} to the
existing layout. There is no formatting step, no coordinator, and no ordering
requirement between engines: whichever starts first formats, the other joins.

\subsection{The directory lives inside the region}

A shared key-value store requires a directory. Instead of an external service
like Redis, etcd, or an RDMA-side registry, we place it in the shared medium
itself (Figure~\ref{fig:layout}(b)).

KV blocks are named by a content hash of the tokens they cover. The CXL layer
takes the first 16 hex characters of vLLM's offload-key basename as a 64-bit
\emph{slot key}. The slot index is \code{key \% num\_slots} with linear
probing. Three arrays in the region implement the directory: a bitmap of
occupancy (\code{atomic<uint64\_t>} words), a parallel array of slot keys, and
a per-slot CRC32 array.

The write order is key-last: the slot is reserved first, data is written and 
flushed, the CRC is computed and flushed, and only then is the key published. 
A remote reader only accepts a slot once data and CRC are already on the 
device. The read order invalidates metadata before probing, then verifies 
the CRC before accepting the data.

The CRC array adds physical integrity: it catches bit flips, partial writes, 
and coherence edge cases that a key match alone cannot detect.

Because the directory is just three arrays at known offsets, the
scheduler process answers lookups by \code{mmap}ing the region from Python and
reading them. No metadata service exists to deploy, scale, or fail.

\subsection{The data path, and what it costs}

Figure~\ref{fig:layout}(c) shows both directions. A store copies GPU blocks
into a thread-local host staging buffer, synchronizes the CUDA stream,
reserves a slot, \code{memcpy}s to the CXL region, flushes data then CRC, and
publishes the key last. A load invalidates metadata, probes for the slot,
verifies the CRC, \code{memcpy}s from the region into staging, and copies
staging to the GPU.

This is \textbf{host-staged, not zero-copy}: the GPU never DMAs to or from CXL.
\S\ref{sec:bandwidth} quantifies the cost: we achieve 5.8\,GB/s against a
27\,GB/s fabric, leaving 3--5$\times$ headroom unrealized. Zero-copy will
require future CXL devices with peer-to-peer (P2P) support.

We also have no eviction in the prototype. Slots are effectively
write-once: no TTL, no LRU, no
background reclaim. A store that finds no free slot fails gracefully.

\subsection{Coherence}

Correctness across nodes rests on software managed CXL.mem coherence over
the DAX mapping. Our
experiments provide strong \emph{behavioral} evidence: tens of thousands of
cross-node hits, a zero-false-positive control run, a same-versus-cross
latency difference of 1--4\,\%, and zero CRC32 mismatches over 1.2\,TB of
writes (\S\ref{sec:c0}), but we did not verify loaded bytes against written
bytes directly, and \S\ref{sec:c0} explains why our attempt was
inconclusive.

\subsection{Implementation notes}

The KV tier is a C++ engine with Python glue behind vLLM's
\code{OffloadingConnector}, running inside vLLM v0.23.0. During development, a
seeding issue invalidated an early dataset: vLLM 0.23 seeds its
block-hash chain from \code{os.urandom()} when \code{PYTHONHASHSEED} is unset.
Two engines therefore derive \emph{different} hashes for identical tokens, and
every cross-instance lookup misses silently. \code{PYTHONHASHSEED} must be set
to the same value on every replica sharing a region.

\section{Evaluating Shared Memory for Cross-Node KV Reuse}
\label{sec:eval}

All values in this section are computed by our analysis script from the raw
per-request logs; every table and figure is generated from the same
\code{stats.json}.

\subsection{Testbed}

\paragraph{Compute nodes.} Two worker nodes, each a single-socket AMD EPYC
9254 (24c/48t), 128\,GB DDR5, one NVIDIA L4 (24\,GB), running Kubernetes
v1.35.7. The nodes are directly connected by 100\,GbE ConnectX-6 Dx RoCEv2,
characterized in \S\ref{sec:fabric}.

\paragraph{CXL appliance.} An FPGA-based prototype memory blade for the Seagate
Composable Memory Appliance. It provides two host ports, each running CXL\,1.1
over PCIe\,Gen5\,x16, and holds 1\,TiB of composable resource backed by
8$\times$128\,GiB DDR4 modules at 1866\,MHz in 2DPC. The two-port topology
yields a two-node cluster: each port connects to one host, and a composed region
is multi-assigned to both. We compose 512\,GiB, rather than the full 1\,TiB, to
avoid an AMD BIOS memory hole at the 1\,TiB address boundary. A production
appliance with more ports would let the same driver serve four or more consumers
on a single blade.

\paragraph{Software and model.} Serving uses vLLM v0.23.0 with
\code{llm-d}, one replica per node at tensor parallelism 1. The model is
Qwen2.5-7B-Instruct~\cite{qwen25} (bf16, 56\,KiB of KV per token).

\paragraph{GPU capacity.} A single NVIDIA L4 (24\,GB VRAM) per node is
sufficient for closed-loop, one-session-at-a-time prefill measurements but
prevents us from running concurrent decode traffic alongside the reusing
request: a 32\,K-token prefill plus active decode slots would exceed VRAM
capacity. Consequently we cannot report goodput, tail latency under load, or
contention behavior: \S\ref{sec:limits} discusses the implications.

Table~\ref{tab:testbed} summarizes the testbed and its measured
characterization.

\begin{table}[t]\centering\small
\caption{Testbed and measured tier characterization. Latency and bandwidth are measured with Intel MLC on both worker nodes.}\label{tab:testbed}
\begin{tabular}{@{}lll@{}}
\toprule
Tier & Latency & Bandwidth \\
\midrule
Local DRAM (NUMA 0) & 111\,ns & 134\,GB/s \\
CXL appliance (512\,GiB, dual-attached) & 461--507\,ns & 27.1--28.5\,GB/s \\
RoCEv2 (100\,GbE, direct-connect) & 1.13\,$\mu$s & 12.25\,GB/s \\
\midrule
Worker nodes & 2$\times$ EPYC 9254, 128\,GB DDR5, 1$\times$ L4 & \\
Software & K8s v1.35.7, vLLM v0.23.0, \texttt{llm-d} & \\
Model & Qwen2.5-7B-Instruct, bf16, 56\,KiB/token & \\
\bottomrule
\end{tabular}\end{table}

\subsection{What the fabric can do}
\label{sec:fabric}

Memory Latency Checker~\cite{mlc} measurements on both nodes report: local DRAM
at 111\,ns / 134\,GB/s; the CXL appliance at 461--507\,ns / 27.1--28.5\,GB/s;
and the RoCEv2 link at 1.13\,$\mu$s / 12.25\,GB/s (98\,\% of line rate).
These numbers are consistent with prior CXL performance studies on data
workloads~\cite{cxl-vldb2025} and shared-memory access
benchmarks~\cite{cxl-bench-adms2025}. On this hardware CXL is 2.4$\times$ lower
latency and 2.2$\times$ higher bandwidth than the RDMA fabric, and
byte-addressable: a KV block is read with ordinary loads rather than a
transfer protocol.

\subsection{Methodology}
\label{sec:method}

\paragraph{What we measure.} Client-observed TTFT over SSE streaming, from the
first byte of the request to the first generated token. Hit rates come from
vLLM's \code{external\_prefix\_cache\_\{queries,hits\}} counters, scraped on
\textbf{both} replicas before and after \textbf{each} request.

\paragraph{The experiment.} Each session issues two requests sharing a long
prefix:

\begin{center}\small
\begin{tabular}{@{}ll@{}}
\code{req1 = SHARED + Q1} & $\rightarrow$ replica A \quad(writes KV into the
tier)\\
\code{req2 = SHARED + Q2} & $\rightarrow$ replica A (\emph{same}) \ $|$ replica
B (\emph{cross})\\
\end{tabular}
\end{center}

The \textbf{cross} arm isolates the shared-memory effect: replica B never saw
those tokens, and node-local tiers (GPU prefix caching, CPU-DRAM offload)
cannot serve it. Placement is
deterministic because we drive the replicas directly. The cross arm also
avoids a short-circuit in the connector's manager, which records locally stored
keys and answers same-replica lookups without consulting the region.

\paragraph{Grid.} 5 tiers (Table~\ref{tab:configs}) $\times$ 2 arms $\times$ 4
prefix lengths $\times$ 20 sessions $\times$ 3 repetitions = 2\,400 request
pairs. Prefix lengths are tokenizer-calibrated; the nominal
2\,K/8\,K/16\,K/32\,K buckets are medians of 2\,054 / 8\,173 / 16\,282 /
32\,694 tokens. Body text is generated per session from a seeded template
grammar. Tier order is varied per repetition so that ordering and thermal
drift cannot confound the tier effect, and the CXL region is wiped between
repetitions.

\begin{table}[t]\centering\small
\caption{The five KV-reuse configurations under comparison. T3 and T4 place KV blocks somewhere a replica on the other node can reach.}\label{tab:configs}
\begin{tabular}{@{}llp{0.34\linewidth}c@{}}
\toprule
Tier & KV reuse mechanism & vLLM configuration & Shared? \\
\midrule
T0 & none (recompute floor) & \texttt{-{}-no-enable-prefix-caching} & --- \\
T1 & GPU VRAM prefix cache & \texttt{-{}-enable-prefix-caching} & no \\
T2 & CPU DRAM offload & \texttt{-{}-no-enable-prefix-caching} \texttt{-{}-kv-offloading-backend=native} \texttt{-{}-kv\_offloading\_size=32} & no \\
T3 & RDMA(Mooncake Store) & \texttt{OffloadingConnector}, \texttt{backend=mooncake}, \texttt{-{}-no-enable-prefix-caching} & yes \\
T4 & \textbf{shared CXL region (ours)} & \texttt{OffloadingConnector}, \texttt{backend=CXL}, \texttt{cxl\_mode=read\_write}, \texttt{-{}-no-enable-prefix-caching} & \textbf{yes} \\
\bottomrule
\end{tabular}\end{table}

\paragraph{Prerequisites.} \code{PYTHONHASHSEED=0} on both replicas
(\S\ref{sec:connector}), and an offload granularity of 256 tokens, so any prefix
not a multiple of 256 recomputes its trailing partial block on every tier.

\paragraph{Validity controls.} Three checks run on every cell.

\begin{enumerate}
\item \textbf{No prompt leakage.} \code{req1} hit rate must be $\approx$0.
  Measured: $\le$0.2\,\% on every tier; 32 of 32 cells pass a 5\,\% threshold.
\item \textbf{No false hits.} A separate control run confirms \textbf{0 hits
  out of 12\,298 queries} for never-stored keys, ruling out the bitmap-only
  defect of \S\ref{sec:connector}.
\item \textbf{Fresh region.} The harness aborts a T4 run unless exactly one
  replica logs an initialize and the other a reuse.
\end{enumerate}

\paragraph{Bounding the same-versus-cross bias.} The two nodes are not perfectly
matched, and each arm has its own prompt set. T0 measures both at once: its
cross-arm median TTFT is 1.4\%, 3.6\%, 4.5\%, and 4.6\% \emph{below} its
same-arm median at 2\,K, 8\,K, 16\,K, and 32\,K. Because the cross arm is
naturally faster (by up to $\sim$5\,\%), the collapse of T1/T2 in the cross arm
is conservative: had the bias gone the other way, the penalty would appear
smaller. Thus \textbf{every cross/same ratio carries a combined systematic bias
of $\le$5\,\%}. Tier-to-tier comparisons within the cross arm are unaffected.

\subsection{Only a shared tier delivers cross-node reuse}
\label{sec:c1}

\begin{figure}[t!]\centering
\includegraphics[width=0.8\linewidth]{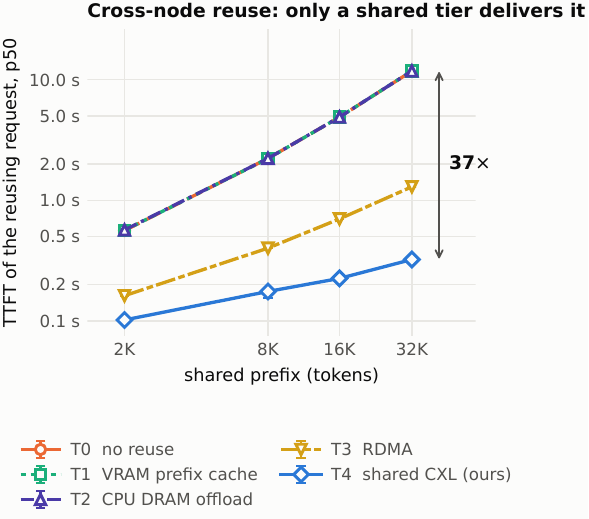}
\caption{Cross-node reuse. When the reusing request lands on the other node, the
GPU prefix cache (T1) and CPU-DRAM offload (T2) coincide with full recompute
(T0): neither cache exists on the node that must serve the request. The Mooncake
Store connector RDMA tier (T3), and the shared CXL tier (T4) both serve it; the
advantage grows with context length. Error bars span the three repetitions.}
\label{fig:c1}
\end{figure}

\begin{table}[t]\centering\small
\caption{Cross-node KV reuse. Speedups are ratios of median TTFT; \emph{cross/same} is 1.0 for a tier that shares perfectly and grows without bound for a node-local one.}\label{tab:c1}
\begin{tabular}{rrrrrrrrr}
\toprule
Prefix & T0 & T1 & T2 & T3 & T4 & T4 vs T0 & T4 hit & \multicolumn{1}{c}{cross/same} \\
 & \multicolumn{5}{c}{median TTFT (ms)} & & & T1 \quad T4 \\
\midrule
2K & 560.7 & 563.3 & 566.8 & 161.5 & \textbf{101.6} & 5.5$\times$ & 95.4\% & 6.5$\times$ \quad 1.02$\times$ \\
8K & 2223.6 & 2218.6 & 2224.6 & 400.7 & \textbf{175.0} & 12.7$\times$ & 98.0\% & 19.0$\times$ \quad 1.04$\times$ \\
16K & 4895.0 & 4898.4 & 4901.9 & 700.4 & \textbf{224.6} & 21.8$\times$ & 99.0\% & 37.2$\times$ \quad 1.04$\times$ \\
32K & 11816.9 & 11845.0 & 11822.3 & 1294.3 & \textbf{322.6} & 36.6$\times$ & 99.5\% & 62.6$\times$ \quad 1.01$\times$ \\
\bottomrule
\end{tabular}\end{table}

Figure~\ref{fig:c1} and Table~\ref{tab:c1} give the primary result
(raw TTFT percentiles in Table~\ref{tab:ttft}). In the
cross arm, T1 and T2 are indistinguishable from having no cache at all. 
T3 serves the cross request at 162--1\,294\,ms
median TTFT, with a cross/same penalty of 1.8--6.4$\times$ that grows with
prefix length. The shared CXL
tier (T4) serves it faster at 102--323\,ms, yielding a 5.5--36.6$\times$
speedup over recompute and a 1.6--4.0$\times$ speedup over T3 (median TTFT,
$n=60$ per cell).

The speedup grows with prefix length because the two sides scale differently.
Recompute is superlinear in context, with an effective prefill rate that falls
from 3\,657 to 2\,767\,tok/s between 2\,K and 32\,K, while the CXL read path is
dominated by data movement. A least-squares fit gives
\begin{equation}
\mathrm{TTFT} \approx 104\,\mathrm{ms} + 6.90\,\mu\mathrm{s/token}
\label{eq:fit}
\end{equation}
with residuals $<$17\,ms.

Reuse has a break-even prefix length below which the tier's fixed
cost is not worth paying. Extrapolating the two trends puts it in the
\emph{low hundreds of tokens}: below this length, the CXL tier's
$\sim$104\,ms fixed overhead exceeds the recompute cost, so the tier
should not be consulted.

\subsection{TTFT penalty across nodes}
\label{sec:gap}

\begin{figure}[t!]\centering
\includegraphics[width=0.9\linewidth]{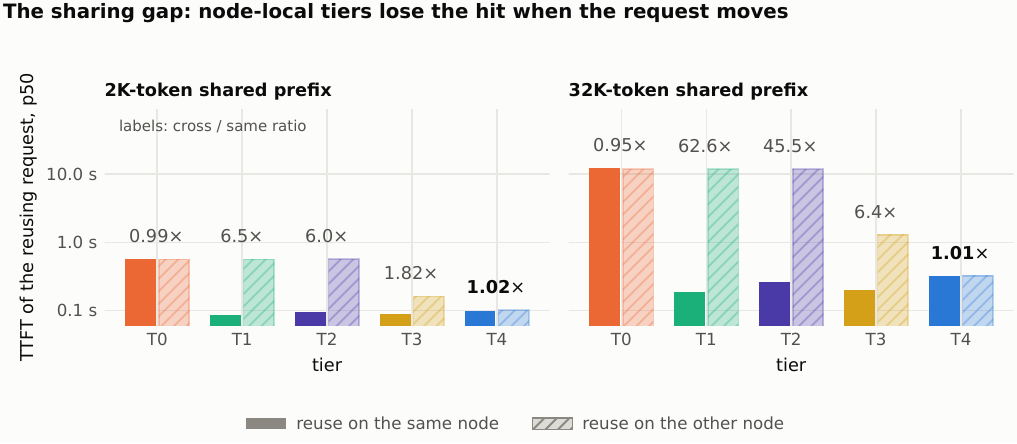}
\caption{The sharing gap: the same data read as a within-tier question. What
does it cost a tier when the reusing request moves to the other node? For the
node-local tiers the hit is simply lost; for the shared region there is no
boundary to cross.}
\label{fig:gap}
\end{figure}

Figure~\ref{fig:gap} asks the within-tier question: \emph{what does it cost a
tier when the reusing request moves?}

\begin{table}[tbp]\centering\small
\caption{The sharing gap: cross-arm median TTFT divided by same-arm median TTFT. A value of 1.0 means the tier shares perfectly; node-local tiers lose their hit entirely when the request moves.}\label{tab:sharing-gap}
\begin{tabular}{@{}lrrrrr@{}}
\toprule
Prefix & T0 & T1 & T2 & T3 & T4 \\
\midrule
2K & 0.99$\times$ & 6.5$\times$ & 6.0$\times$ & 1.8$\times$ & \textbf{1.02$\times$} \\
8K & 0.96$\times$ & 19.0$\times$ & 16.1$\times$ & 3.3$\times$ & \textbf{1.04$\times$} \\
16K & 0.95$\times$ & 37.2$\times$ & 28.8$\times$ & 5.0$\times$ & \textbf{1.04$\times$} \\
32K & 0.95$\times$ & 62.6$\times$ & 45.5$\times$ & 6.4$\times$ & \textbf{1.01$\times$} \\
\bottomrule
\end{tabular}\end{table}

Table~\ref{tab:sharing-gap} summarizes the ratio (raw medians from
Table~\ref{tab:ttft}). T1 and T2 lose their hit entirely when the request moves:
a 6--63$\times$ penalty growing with context. \textbf{T4 is within 1--4\,\% of
its own same-node number at every prefix length.} A 36$\times$ speedup
quantifies the recompute penalty; a 1.02$\times$ cross/same ratio quantifies the
sharing cost.

\subsection{Where the hits and the time go}
\label{sec:bandwidth}

\begin{figure}[t!]\centering
\includegraphics[width=\linewidth]{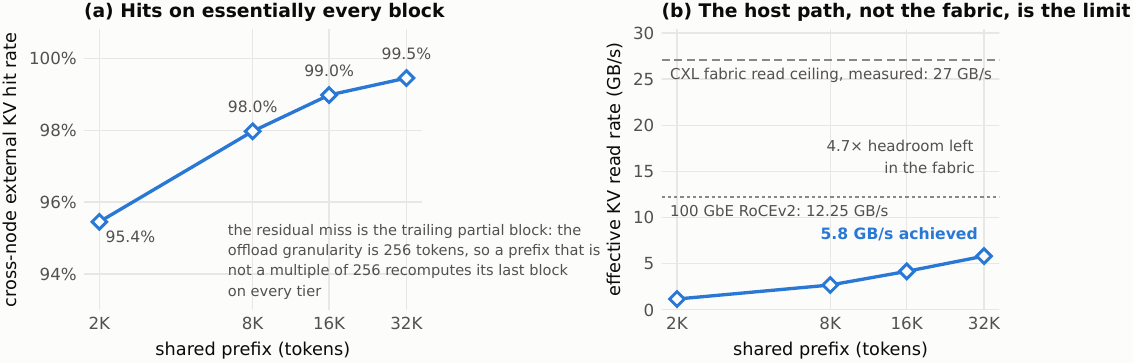}
\caption{(a) Cross-node external hit rate against the 256-token
block-alignment bound: every fully aligned block hits, and the residual is one
trailing partial block. (b) Achieved effective read rate against the measured
fabric ceiling: the limit we measured is our host-staged copy, not CXL.}
\label{fig:hits}
\end{figure}

\begin{table}[t]\centering\small
\caption{Effective KV read rate achieved by the shared tier, computed as the reloaded KV volume divided by the whole median TTFT (so it includes the engine's fixed cost). The fabric ceiling is the 27.1\,GB/s measured in Table~\ref{tab:testbed}.}\label{tab:bw}
\begin{tabular}{@{}lrrrr@{}}
\toprule
Prefix & KV volume & T4 p50 TTFT & effective read & \% of ceiling \\
\midrule
2K & 112\,MiB & 101.6\,ms & 1.16\,GB/s & 4\% \\
8K & 447\,MiB & 175.0\,ms & 2.68\,GB/s & 10\% \\
16K & 891\,MiB & 224.6\,ms & 4.16\,GB/s & 15\% \\
32K & 1788\,MiB & 322.6\,ms & 5.81\,GB/s & 21\% \\
\bottomrule
\end{tabular}\end{table}

\paragraph{Hit rate.} Cross-node external hit rate is 95.4\,\% / 98.0\,\% /
99.0\,\% / 99.5\,\% at 2\,K/8\,K/16\,K/32\,K (Figure~\ref{fig:hits}(a)). This is
the \textbf{block-alignment bound}: the offload granularity is 256 tokens, the
trailing partial block is recomputed on every tier, and measured hits track the
aligned fraction to within 0.9\,pp at $\ge$8\,K (and 0.05\,pp at $\ge$16\,K).
Every fully aligned block hits.

\paragraph{Where the time goes.} Equation~\ref{eq:fit} decomposes the
cross-arm TTFT into a fixed component and a per-token marginal cost:

\begin{itemize}
\item \textbf{$\sim$104\,ms fixed.} Most of this is engine overhead, not CXL:
  T1 (a pure VRAM hit, same arm) costs 87\,ms at 2\,K, so the CXL path adds
  only $\sim$13\,ms on top of the vLLM floor (CUDA graph setup, scheduler
  dispatch, Python glue). The intercept is not a CXL overhead: it is
  the cost of running the engine.
\item \textbf{6.90\,$\mu$s/token marginal}, i.e.\ 8.3\,GB/s on the data
  path, or 31\,\% of the 27\,GB/s fabric bandwidth limit.
\end{itemize}

The bottleneck is the host-staged data path: KV is copied CXL $\rightarrow$ CPU
staging $\rightarrow$ GPU (two \code{memcpy} hops) rather than DMA'd.
The measured rate is limited by this implementation choice; a zero-copy path
offers 3.3$\times$ headroom on the marginal rate.

Dividing the reloaded KV volume by the full TTFT (fixed + marginal) yields a
lower effective rate because the $\sim$104\,ms fixed cost amortizes over the
transfer. At 32\,K tokens the effective rate is 5.81\,GB/s (21\,\% of the limit;
Table~\ref{tab:bw}, Figure~\ref{fig:hits}(b)), rising toward the marginal
8.3\,GB/s as prefix length grows. At short prefixes the effective rate is
low (1.16\,GB/s at 2\,K) because the fabric is idle during the
engine's fixed overhead, not saturated during the copy.

\subsection{TTFT distribution and first-touch warm-up}

\begin{figure}[t!]\centering
\includegraphics[width=\linewidth]{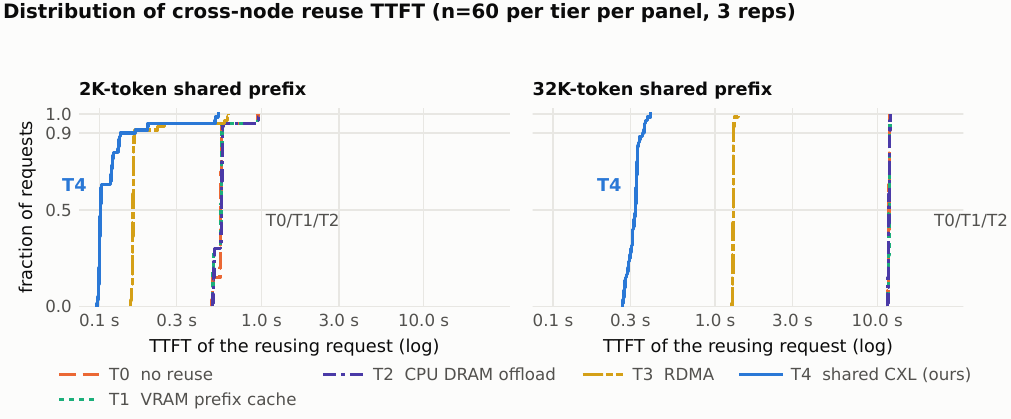}
\caption{TTFT distribution in the cross arm at the shortest and longest prefix
($n=60$ per tier per panel). Distributions are tight for every tier; this is a
closed-loop, one-session-at-a-time harness, so no tail-latency-under-load claim
can be made from it. T4's only visible tail is the single first-touch request
discussed in the text.}
\label{fig:cdf}
\end{figure}

\begin{figure}[t!]\centering
\includegraphics[width=0.85\linewidth]{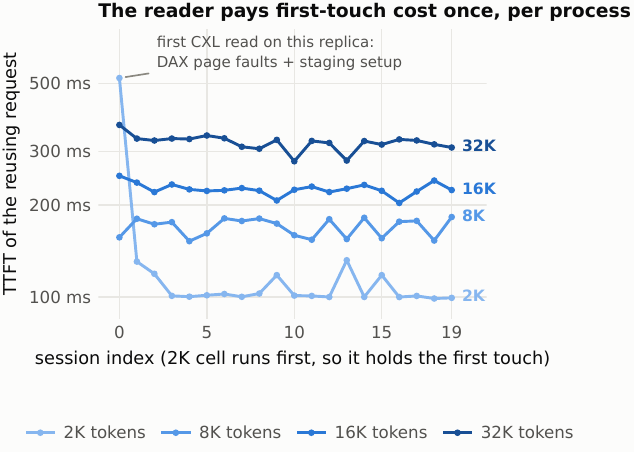}
\caption{The reading replica pays a $\sim$420\,ms first-touch cost once per
process, from DAX page faults and staging-buffer allocation, and never again.
The 2\,K cell runs first and absorbs it; the longer cells that follow show no
spike.}
\label{fig:warmup}
\end{figure}

\begin{table}[t]\centering\small
\caption{TTFT of the reusing request (ms), $n=60$ per cell (3 repetitions $\times$ 20 sessions). \emph{cross} means the second request is served by the replica on the other node, so only a shared tier can hit. Hit\% is the external-only rate.}\label{tab:ttft}
\begin{tabular}{llrrrrrr}
\toprule
Tier & Arm & Prefix & p50 & p90 & p99 & p99/p50 & Hit\% \\
\midrule
T0 & same & 2K & 568.6 & 576.2 & 580.2 & 1.02 & 0.0 \\
T0 & same & 8K & 2306.3 & 2386.1 & 2391.6 & 1.04 & 0.0 \\
T0 & same & 16K & 5127.6 & 5191.8 & 5205.8 & 1.02 & 0.0 \\
T0 & same & 32K & 12388.8 & 12504.8 & 12585.7 & 1.02 & 0.0 \\
T0 & cross & 2K & 560.7 & 568.7 & 949.7 & 1.69 & 0.0 \\
T0 & cross & 8K & 2223.6 & 2279.6 & 2284.8 & 1.03 & 0.0 \\
T0 & cross & 16K & 4895.0 & 4950.5 & 4966.3 & 1.01 & 0.0 \\
T0 & cross & 32K & 11816.9 & 11924.7 & 11979.6 & 1.01 & 0.0 \\
\midrule
T1 & same & 2K & 86.6 & 87.3 & 88.0 & 1.02 & 0.0 \\
T1 & same & 8K & 117.0 & 120.8 & 122.1 & 1.04 & 0.0 \\
T1 & same & 16K & 131.8 & 148.6 & 149.8 & 1.14 & 0.0 \\
T1 & same & 32K & 189.1 & 191.0 & 192.0 & 1.02 & 0.0 \\
T1 & cross & 2K & 563.3 & 570.7 & 958.6 & 1.70 & 0.0 \\
T1 & cross & 8K & 2218.6 & 2287.1 & 2296.7 & 1.04 & 0.0 \\
T1 & cross & 16K & 4898.4 & 4937.4 & 4959.4 & 1.01 & 0.0 \\
T1 & cross & 32K & 11845.0 & 11921.1 & 12015.7 & 1.01 & 0.0 \\
\midrule
T2 & same & 2K & 93.9 & 94.9 & 95.5 & 1.02 & 98.8 \\
T2 & same & 8K & 138.4 & 142.5 & 143.4 & 1.04 & 99.7 \\
T2 & same & 16K & 170.3 & 187.1 & 189.4 & 1.11 & 99.9 \\
T2 & same & 32K & 259.9 & 265.0 & 267.7 & 1.03 & 99.9 \\
T2 & cross & 2K & 566.8 & 575.7 & 954.8 & 1.68 & 0.7 \\
T2 & cross & 8K & 2224.6 & 2280.9 & 2298.9 & 1.03 & 0.4 \\
T2 & cross & 16K & 4901.9 & 4944.4 & 4962.8 & 1.01 & 0.2 \\
T2 & cross & 32K & 11822.3 & 11926.3 & 12023.7 & 1.02 & 0.1 \\
\midrule
T3 & same & 2K & 88.5 & 89.2 & 90.9 & 1.03 & 0.0 \\
T3 & same & 8K & 122.7 & 125.7 & 126.2 & 1.03 & 0.0 \\
T3 & same & 16K & 139.4 & 154.8 & 158.1 & 1.13 & 0.0 \\
T3 & same & 32K & 203.7 & 208.6 & 209.8 & 1.03 & 0.0 \\
T3 & cross & 2K & 161.5 & 164.7 & 615.9 & 3.81 & 98.9 \\
T3 & cross & 8K & 400.7 & 403.5 & 406.0 & 1.01 & 99.7 \\
T3 & cross & 16K & 700.4 & 708.2 & 715.0 & 1.02 & 99.9 \\
T3 & cross & 32K & 1294.3 & 1300.2 & 1311.7 & 1.01 & 99.9 \\
\midrule
T4 & same & 2K & 99.5 & 130.1 & 131.8 & 1.32 & 94.6 \\
T4 & same & 8K & 168.4 & 178.8 & 195.9 & 1.16 & 97.9 \\
T4 & same & 16K & 216.3 & 234.6 & 239.7 & 1.11 & 98.9 \\
T4 & same & 32K & 320.6 & 347.7 & 366.9 & 1.14 & 99.5 \\
T4 & cross & 2K & 101.6 & 134.7 & 521.2 & 5.13 & 95.4 \\
T4 & cross & 8K & 175.0 & 196.2 & 199.0 & 1.14 & 98.0 \\
T4 & cross & 16K & 224.6 & 240.6 & 249.3 & 1.11 & 99.0 \\
T4 & cross & 32K & 322.6 & 357.4 & 381.4 & 1.18 & 99.5 \\
\bottomrule
\end{tabular}\end{table}

TTFT distributions are tight for every tier (Figure~\ref{fig:cdf},
Table~\ref{tab:ttft}). At $\ge$8\,K tokens T4's p99/p50 is 1.11--1.18,
comparable to T0's 1.01--1.03 while sitting an order of magnitude lower in
absolute terms.

T4's only tail is at 2\,K tokens (p99/p50 = 5.13), and it is one request: the
reading replica's first CXL read costs \textbf{$\sim$420\,ms extra}
(521\,ms vs.\ 101\,ms steady state) for DAX first-touch page faults and
staging-buffer allocation. It is paid once per process: the 8\,K, 16\,K and
32\,K cells that follow show no spike (Figure~\ref{fig:warmup}). In the whole
cross arm exactly \textbf{3 of 240 requests} exceed 450\,ms, one per
repetition, all first reads.

\subsection{Cross-node availability beyond VRAM capacity}

\begin{table}[t]\centering\small
\caption{The honest cost of the tier when reuse \emph{is} local (same arm): the GPU's own prefix cache is faster than the shared region, by at most 1.7$\times$.}\label{tab:local}
\begin{tabular}{@{}lrrr@{}}
\toprule
Prefix & T1 p50 (ms) & T4 p50 (ms) & T4/T1 \\
\midrule
2K & 86.6 & 99.5 & 1.15$\times$ \\
8K & 117.0 & 168.4 & 1.44$\times$ \\
16K & 131.8 & 216.3 & 1.64$\times$ \\
32K & 189.1 & 320.6 & 1.70$\times$ \\
\bottomrule
\end{tabular}\end{table}

When reuse \emph{is} local, VRAM prefix caching still wins by
1.15$\times$--1.70$\times$ (Table~\ref{tab:local}) because no remote tier
can match the bandwidth of memory on the same die as the compute.

A shared CXL region is visible to every attached host simultaneously and
persists independently of per-replica VRAM pressure. Consider a popular
prefix evicted from one GPU's prefix cache under memory pressure: the CXL
tier still serves it to any replica that needs it, avoiding a full
re-prefill. The tier complements prefix caching by serving requests when
the per-GPU budget is exceeded or the request has been load-balanced to a
different pod. The trade:
\textbf{$\le$1.7$\times$ premium when reuse is local, against a 6--63$\times$
penalty when it is not.}

\subsection{Prefill/decode disaggregation}
\label{sec:pd}

The primary experiment demonstrates cross-node prefix cache reuse: both
replicas are full engines. We extend the evaluation with prefill/decode (P/D)
disaggregation, where the prefill and decode phases run on separate pods and
the KV cache is transferred between them. We run two configurations:
(i) P/D over the shared CXL region (the prefill pod writes KV to CXL, the
decode pod reads it from CXL), and (ii) P/D over RDMA (the prefill
pod sends KV to the decode pod via the Mooncake connector over RoCEv2). 
We measure only the TTFT of the original query, which includes KV-cache
transfer through the selected protocol between the two nodes. Prefix
cache reuse is not evaluated in this section.

\paragraph{P/D over CXL.} The prefill pod runs with \code{kv\_role=kv\_producer}
and the decode pod with \code{kv\_role=kv\_consumer}; both use the
\code{OffloadingConnector} with \code{backend=CXL}. The prefill pod writes the
KV blocks for the prompt into the shared CXL region; the decode pod reads them
from CXL before starting generation. Median TTFT: 758\,ms / 2\,696\,ms /
5\,693\,ms / 13\,275\,ms at 2\,K/8\,K/16\,K/32\,K, with $\le$2.5\,\% rep-to-rep
spread.

\paragraph{P/D over RDMA.} The prefill and decode pods use the Mooncake
connector over RDMA for cross comparison; the KV cache is transferred over
RoCEv2. Median TTFT: 711\,ms / 2\,623\,ms / 5\,639\,ms / 13\,298\,ms at
2\,K/8\,K/16\,K/32\,K, with tight distributions (p50$\to$p99 spread
1--2\,\%).

\begin{figure}[t]
\centering
\includegraphics[width=0.8\linewidth]{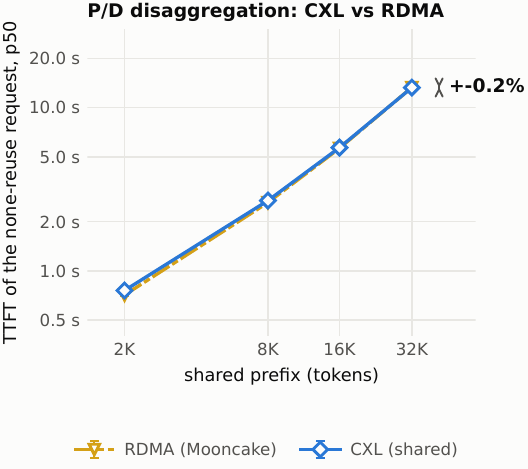}
\caption{P/D disaggregation: TTFT of the none-reuse request, over CXL vs.\ RDMA.
Error bars show the interquartile range (p25--p75) across 60 samples (3 reps
$\times$ 20 sessions). The two paths converge to within 7\,\% at all prefix
lengths.}
\label{fig:pd}
\end{figure}

\paragraph{Comparison.} The P/D CXL and RDMA paths are within 7\,\% of each
other at every prefix length (Figure~\ref{fig:pd}): median TTFT differs by
6.6\,\% at 2\,K
(758\,ms vs.\ 711\,ms), 2.8\,\% at 8\,K (2\,696\,ms vs.\ 2\,623\,ms),
1.0\,\% at 16\,K (5\,693\,ms vs.\ 5\,639\,ms), and 0.2\,\% at 32\,K
(13\,275\,ms vs.\ 13\,298\,ms). TTFT is dominated by
GPU compute; KV transfer accounts for a small fraction of the total time.
This experiment shows that CXL shared memory is comparable in performance to
the standard RDMA-based approach in the P/D disaggregated setup.

\subsection{Correctness: end-to-end integrity with bf16 rounding}
\label{sec:c0}

This experiment runs on CXL 1.1 on both hosts and the multi-headed shared memory
device. There is no hardware-based coherency or write ordering enforced between
the two host ports. We implemented cache-line eviction and invalidation-based
software coherency management in the design and cross-node integrity checks.
Cross-node integrity checking relies on three layers: vLLM's content-addressable
block hash (same tokens produce the same block identity), the per-slot key array
(eliminates false-positive reads from hash collisions), and a per-slot CRC32
checksum over the exact byte range written (catches physical corruption). A full
cross-replica benchmark reports \textbf{zero CRC32 mismatches} over
approximately 1.2\,TB of CXL writes, with a 95\,--\,99\,\% external hit rate on
the reading pod. The cross-replica TTFT matches the same-replica TTFT to within
1\,--\,4\,\% (Table~\ref{tab:sharing-gap}); corrupted data would trigger cache
misses and prefill fallback, making the cross arm significantly slower.

Replaying 10 identical prompts through T0 and T4 (greedy decoding,
\code{temperature=0}, \code{seed=0}) yields \textbf{6/10 byte-identical
continuations}. The four divergences are caused by bf16 rounding: near-tie
values may round differently across GPUs, producing at most a 1\,ULP difference
in the mantissa. The CRC32 protects against \emph{corruption}, not
\emph{divergence}: a legitimate rounding difference produces a valid checksum.
Because the KV cache is an intermediate representation, a 1\,ULP difference in a
bf16 attention score has negligible impact on generation quality in practice.

\subsection{Reproducibility}

\begin{table}[t]\centering\small
\caption{Reproducibility: per-repetition median TTFT (ms) in the cross arm. \emph{spread} is (max$-$min)/median. Repetitions ran in different tier orders on different days.}\label{tab:repro}
\begin{tabular}{@{}llrrrr@{}}
\toprule
Tier & Prefix & rep 1 & rep 2 & rep 3 & spread \\
\midrule
T0 & 2K & 559.4 & 560.1 & 566.5 & 1.3\% \\
T0 & 8K & 2221.0 & 2236.4 & 2255.2 & 1.5\% \\
T0 & 16K & 4889.9 & 4891.0 & 4900.3 & 0.2\% \\
T0 & 32K & 11798.4 & 11821.7 & 11826.9 & 0.2\% \\
\midrule
T1 & 2K & 562.2 & 564.1 & 565.2 & 0.5\% \\
T1 & 8K & 2214.8 & 2218.8 & 2219.9 & 0.2\% \\
T1 & 16K & 4895.7 & 4898.4 & 4904.4 & 0.2\% \\
T1 & 32K & 11843.9 & 11847.4 & 11852.9 & 0.1\% \\
\midrule
T2 & 2K & 564.8 & 566.6 & 570.0 & 0.9\% \\
T2 & 8K & 2209.9 & 2224.0 & 2228.3 & 0.8\% \\
T2 & 16K & 4899.2 & 4901.3 & 4903.7 & 0.1\% \\
T2 & 32K & 11801.3 & 11810.1 & 11840.9 & 0.3\% \\
\midrule
T3 & 2K & 161.0 & 161.5 & 161.9 & 0.6\% \\
T3 & 8K & 399.1 & 400.7 & 401.0 & 0.5\% \\
T3 & 16K & 699.8 & 700.2 & 701.6 & 0.3\% \\
T3 & 32K & 1288.2 & 1292.5 & 1297.0 & 0.7\% \\
\midrule
T4 & 2K & 101.1 & 101.8 & 101.9 & 0.8\% \\
T4 & 8K & 155.7 & 173.6 & 178.1 & 12.9\% \\
T4 & 16K & 223.7 & 224.2 & 227.7 & 1.8\% \\
T4 & 32K & 315.4 & 317.8 & 325.3 & 3.1\% \\
\bottomrule
\end{tabular}\end{table}

Three full repetitions in different tier orders. Per-repetition median TTFT
(Table~\ref{tab:repro}) agrees to within 3.1\,\% on 15 of 16 cross-arm cells;
the exception is T4 at 8\,K tokens, where the three medians span 12.9\,\%
(156 / 174 / 178\,ms), still an order of magnitude inside the effect being
measured.

The remaining limitations are collected in \S\ref{sec:limits}.

\section{Related Work}

KV-cache reuse has moved from single-server prefix caching to cross-node
pooled storage. vLLM~\cite{vllm} introduced PagedAttention, which virtualizes
KV memory inside a single GPU; SGLang~\cite{sglang} added aggressive prefix
reuse. Systems like LMCache~\cite{lmcache} and
CacheBlend~\cite{cacheblend} extend the cache to host DRAM and remote
object storage, while Mooncake~\cite{mooncake} pools KV across nodes via
RDMA. Three recent systems target CXL specifically:
TraCT~\cite{tract} demonstrates rack-scale CXL KV transfer for disaggregated
serving, SAC~\cite{sac} targets sparse-attention workloads, and
HyMCache~\cite{hymcache} proposes a CXL memory rack for multi-turn serving.

\begin{table}[t!]\centering\small
\caption{Comparison of KV-cache sharing and CXL memory systems.}\label{tab:rw-comparison}
\begin{tabular}{@{}lllll@{}}
\toprule
System & Medium & K8s DRA & P/D-disagg. & Real CXL hw? \\
\midrule
Mooncake~\cite{mooncake} & RDMA & no & yes & --- \\
TraCT~\cite{tract} & CXL & no & yes & yes \\
HyMCache~\cite{hymcache} & CXL & no & no & yes \\
SAC~\cite{sac} & CXL & no & no & yes \\
Pangaea\,v2~\cite{pangaea2} & CXL & no (NRI) & no & yes \\
\midrule
\rowcolor{gray!5}
\textbf{Ours} & \textbf{CXL} & \textbf{yes} & \textbf{yes} & \textbf{yes} \\
\bottomrule
\end{tabular}
\end{table}

Table~\ref{tab:rw-comparison} contrasts these systems with our approach.
TraCT, SAC, and HyMCache assume the CXL region already exists and focus on
the KV storage layer. Our DRA driver makes the region itself schedulable
by Kubernetes. Mooncake achieves pooled storage via RDMA but requires a
transfer protocol on every access; our CXL tier uses byte-addressable loads
with no transport layer. None of these systems integrate with Kubernetes
scheduling beyond static node assignment.

Prefill/decode disaggregation~\cite{distserve,splitwise} separates phases to
improve accelerator utilization but does not address KV sharing across
replicas. Our memory disaggregation is complementary and can underlie a
P/D handoff.

The broader memory disaggregation literature, including FaRM~\cite{farm},
Infiniswap~\cite{infiniswap}, and GAM~\cite{gam}, exports memory through
explicit transport protocols, typically RDMA. Composable CXL memory exposes
shared memory through load/store semantics, so applications access it with
ordinary loads and stores. CXL pooling systems like Pond~\cite{pond} and
TPP~\cite{tpp} study cloud-scale pooling and OS-level page placement but do not
target Kubernetes scheduling. Octopus~\cite{octopus} explores sparse topology
for CXL memory pods to improve scalability. Wang et al.~\cite{hitchhikers-cxl}
survey programming and optimization techniques for cache-coherent heterogeneous
interconnects including CXL, NVLink-C2C, and AMD Infinity Fabric. Weisgut et
al.~\cite{cxl-vldb2025} study CXL memory performance for in-memory data
processing, and CXL-Bench~\cite{cxl-bench-adms2025} provides a benchmarking
framework for shared CXL memory access. PolarCXLMem~\cite{polarcxlmem}
demonstrates CXL switch-based disaggregated memory for cloud-native databases,
showing up to 2.1$\times$ throughput improvement over RDMA in pooling scenarios.

Kubernetes Dynamic Resource Allocation~\cite{drakep,dra-arch} generalized
the device plugin model with DeviceClasses, ResourceSlices, and
ResourceClaims. Existing DRA drivers target node-local resources (GPUs,
FPGAs, networking). Pangaea\,v2~\cite{pangaea2} brings CXL memory into
Kubernetes but only on a per-node basis. We are not aware of any DRA-managed
memory resource that is dynamically composed and simultaneously attached to
multiple nodes.

\section{Limitations and Threats to Validity}
\label{sec:limits}

\begin{description}[leftmargin=0pt,style=unboxed]
\item[Single GPU, no concurrent load.] Each node has one NVIDIA L4 (24\,GB
  VRAM). A 32\,K-token prefill plus active decode slots almost exceeds VRAM
  capacity, so we cannot run background traffic during measurements. The results
  are isolated-prefill TTFT; they do not bound goodput, tail latency under
  queueing, or contention between prefill and decode. On larger GPUs (A100, H100)
  the tier's behavior under load would be the more revealing test: the CXL read
  competes with decode for the PCIe/GPU memory bus. \item[Two-node evaluation.]
  The appliance prototype has two host ports, yielding a two-node cluster
  (\S\ref{sec:eval}). The DRA driver is not two-node-specific: it enumerates ports
  from blade topology and composes regions for any share count. Evaluating four or
  more consumers on a single blade requires a production appliance with more
  ports. \item[Narrow workload.] One model, one size, 100\,\% prefix reuse by
  construction, no reuse-ratio sweep, no capacity pressure, and no multi-tenant
  scheduling experiment. Multi-tenant scheduling is where the DRA driver's value
  should be measured. \item[Host-staged data path.] 5.8\,GB/s achieved against a
  27\,GB/s fabric; 3--5$\times$ headroom unrealized without GPUDirect-style DMA.
  \item[Correctness: bf16 rounding, not corruption.] Cross-node data integrity is
  verified by per-slot CRC32 (zero mismatches over 1.2\,TB of writes). Six of ten
  prompts produce byte-identical continuations (\S\ref{sec:c0}); the remaining
  four differ by at most 1\,ULP due to bf16 rounding across GPUs. Residual
  hazards: the slot key is not re-verified after the copy (alias possible on
  release-and-reacquire), and the store path silently discards write failures
  (invisible to cross-arm results but corrosive under capacity pressure).
  \item[Coherence is software managed.] Hardware CXL.mem coherence is not
  supported on the test platform. Write ordering cannot be guaranteed between the
  host ports, so one host writes and the other reads the shared CXL memory.
  \item[No eviction.] Write-once slots, no TTL, no reclaim: not a production tier.
  \item[No scheduler-side capacity accounting.] Size travels in annotations;
  advertised capacity refreshes on a 30-second cycle; concurrent claims can race
  the same free capacity. \item[Imperfectly matched nodes.] Bounded at $\le$5\,\%
  on cold prefill (T0), conservative in direction for every reported ratio.
\end{description}

\section{Conclusion and Future Work}

This work demonstrates that dynamically composed, multi-host CXL memory 
can be exposed as a first-class Kubernetes DRA resource and used as a 
shared KV-cache tier for LLM serving. We presented a DRA driver that 
makes composable CXL regions schedulable by Kubernetes: the driver 
advertises blade capacity, composes
regions on demand, attaches them to multiple nodes, isolates allocations
via DAX sub-devices, and releases capacity on claim deletion. We also
presented a KV tier for vLLM/llm-d that embeds its directory inside the
shared region, which removes the dependency on an external metadata
service. On a two-node testbed with a 512\,GiB CXL appliance, cross-node
prefix reuse reduced TTFT by 5.5--36.6$\times$ at a hit rate bounded only
by block alignment, while node-local tiers fell back to full recompute.
The sharing gap was 1.01--1.04$\times$ for the shared region compared to
6.5--62.6$\times$ for node-local tiers when the reusing request moved
nodes.

VRAM prefix caching remains faster by up to 1.7$\times$ when reuse is local; the
CXL tier provides availability when the request lands on a different node and is
1.6--4.0$\times$ faster than a pooled RDMA baseline. The P/D disaggregation
experiment (\S\ref{sec:pd}) shows that the CXL and RDMA paths are within 7\,\%
at all prefix lengths. The remaining limitations are in \S\ref{sec:limits}.

Future work includes: \emph{(i)} multi-tenant scheduling under contention with
scheduler-side capacity accounting; \emph{(ii)} enforced write ordering on the
CXL memory across host ports; and \emph{(iii)} a zero-copy
GPU$\leftrightarrow$CXL DMA path to realize the 3--5$\times$ headroom in the
fabric bandwidth.

\section*{Artifact Availability}

The DRA driver, CXL KV connector, benchmark harness, raw per-request logs for
all three repetitions, and the scripts that generate every table and figure in
\S\ref{sec:eval} will be released on the Seagate GitHub
repository~\cite{stx-github} . Regenerating the evaluation from raw logs is
three commands; no number in this paper is transcribed by hand.


\end{document}